\documentclass[twocolumn,showpacs,preprintnumbers]{revtex4}
\usepackage{amssymb}
\usepackage{graphicx}
\usepackage{dcolumn}
\usepackage{bm}

\input{tcilatex}

\begin{document}

\title{Move set, algorithm, and folding kinetics of Monte Carlo simulations
for lattice polymers }
\author{Yu-Pin Luo, Ming-Chang Huang}
\email{ming@phys.cycu.edu.tw}
\author{Yen-Liang Chou}
\affiliation{Department of Physics, Chung-Yuan Christian University, 32023 Chungli, Taiwan}
\author{Tsong-Ming Liaw}
\affiliation{Computing Centre, Academia Sinica, 11529 Taipei, Taiwan }

\begin{abstract}
The effect of different move sets on the folding kinetics of the Monte Carlo
simulations is analysed based on the conformation-network and the
temperature-dependent folding kinetics. A new scheme of implementing
Metropolis algorithm is given. The new method is shown to satisfy the
detailed balance and converge efficiently towards thermal equilibrium. A new
quantity, employed from the continuous time Monte Carlo method, is
introduced to identify effectively the kinetic traps of foldings.
\end{abstract}

\date{\today}
\maketitle

The problem how proteins fold, in milliseconds to seconds, into unique and
stable structures with definite biological functions has recently become
intriguing to the biophysicians\cite{anfinsen}. The kinetic feature of such
problem amounts to the characteristics of the folding paths. Subject to
this, considerable progress have been achieved through numerical studies of
lattice heteropolymers in two or three dimensions\cite{shakhnovich, chan,
camacho}. Although oversimplified, characteristic features obtained from the
simulations, such as folding funnel\cite{onuchic, bryngelson}, folding
bottleneck\cite{shakhnovich} and kinetic traps\cite{chan, cieplak}, have
provided much insights to the kinetic process. However, there exists some
suspicions about the simulation method\cite{sorenson, collet, hoang} and few
ambiguities in the move sets adopted for the simulations\cite{chan, hoang}.
For the former, question about implementing the kinetic Monte Carlo
algorithm was raised and some implementations were shown to violate the
condition of detailed balance\cite{collet}. For the latter, Chan and Dill%
\cite{chan} and Hoang and Cieplak\cite{hoang} stressed the strong dependence
of the folding landscape on the choice of move sets. Consequently, further
clarification and improvement for the methodology remain essential. In Ref.%
\cite{collet, hoang}, proposals of refining the Monte Carlo algorithm were
discussed. In addition, move steps other than the conventional ones, such as
the snake move \cite{hoang}, have also become of interest.

In this Letter we attempt a systematic analysis for the move sets in
relation to the configuration space and develop a new method of implementing
the Metropolis algorithm \cite{metropolis} appropriate for the problem class
of protein foldings. While the conformational networks appear to be a
natural embedding of the folding paths \cite{amaral,scala}, the move sets as
the sets of generators of connections can correlate with the geometry of
such setting. Subsequently, the identification of the configuration space
enables a straightforward program for the new Monte Carlo implementation. We
demonstrate its reliability as well as accuracy and efficiency by means of
explicit computations. Then, the continuous time Monte Carlo (CTMC) method%
\cite{gillespie1, bortz} equipped with the new schem is used to study the
folding kinetics associated with different move sets. Interestingly, by
invoking CTMC simulations, it is found more general to label the kinetic
traps with the new quantity, called viscosity-factor, in contrast to by the
conventional local energy minimum since entropy effects are also taken into
account. \ \ 

A \ specific model system is used to carry out the study. Based on the
considerations given by Ref. \cite{chen}, we consider the $H$-$P$ modele on
a $2D$ square lattice\cite{dill} with the "protein-like" sequence of $16$
monomers specified as $PHPPPHPPHPPPPHHP$. As the consequence of assigning
the contact energies with $E_{H,H}=-3.3$, $E_{H,P}=-2$ and $E_{P,P}=-1$, the
energy of the native state, shown in the inset of Fig. 1(b), appears to be $%
E_{g}=-13.3$. In addition, the mean energy gap $\Delta E_{g}=1.97$ is
obtained by averaging out all the energy differences between the lowest $10$
excited states and the native one. The specific-heat curve $C_{v}\left(
T\right) $ possesses a single peak, representing the phase transition from
the molten globule to the native state, as well as a shoulder at a higher
temperature, signifying the transition from random coil to molten globule.
Moreover, the folding temperature $T_{f}$, defined as the temperature at
which the probability of the native conformation is $1/2$, is at $T_{f}=0.50$
which is near to the specific-heat peak.

We are concerned with move sets consisting of typical move steps, say the
end flip ($ef$), corner shift ($cs$), crankshaft ($cr$) and the rigid
rotation ($rr$)\cite{chan}. The conventional move set $S_{1}$ only allows
for $ef$, $cs$ and $cr$ based on locality. However, the ergodicity is shown
to fail for $S_{1}$\cite{chan}. To be concrete, in two dimensions the move
set $S_{1}$ prohibits the reaching of one conformation from the other ones
for the folding with $16$ monomers and, considerably, the number of such
conformations rapidly increases for more monomers and/or dimensions. The
problem can be remedied by involving moves of $rr$ type which have been
realised in some simple diffusive motions for groups of monomers\cite{chan,
skolnick}. While $ef$ itself can be viewed as short-scale rigid rotation, an
ergodic move set, say $S_{2}$, can be achieved by simply combining $ef$ with 
$rr$. Finally, involving all the four types, an $S_{3}$ move set is also
known to be ergodic.

Different move sets generate different conformation-networks for which, each
conformation is a node and two nodes are connected by an edge if and only if
the two can be transformed to each other via an elementary move of the move
set\cite{bryngelson, scala, betancourt, huang}. The nodes for the
conformation-network are unambiguously given as long as the ergodicity is
fulfilled. For the case at hand, the number of monomers $M=16$ and the total
node number $K=802,075$. On the other hand, the edge distributions among
nodes remain sensitive to the choice of the move sets. Heuristically, few
measures for the edge distributions are shown to provide the global
properties of the conformation networks as folding in the high temperature
limit. \ 

Define the number of edges $k_{i}$ associated with any node $i$ by means of
one elementary move. The corresponding mean values per node $\left\langle
k\right\rangle $ are $9.2$, $20.7$ and $26.3$ for $S_{1}$, $S_{2}$ and $%
S_{3} $, respectively. The $\left\langle k\right\rangle $ value for $S_{2}$
or $S_{3}$ is twice more than the value for $S_{1}$. Thus, the larger $%
\left\langle k\right\rangle $ value provides more through-way accessibility
to the native state and reduce the possibility of being trapped in local
minimum. The minimal edge number $l_{i}$ for a node $i$ to connect with the
native state is also of interest. Upon enumerating this for the model
sequence with respect to various move sets and taking average, we obtain $%
\left\langle l\right\rangle =\ 26.4$ for $S_{1}$ and $8.4$ for both $S_{2}$
and $S_{3}$. The $\left\langle l\right\rangle $ value for $S_{2}$ or $S_{3}$
is about one-third of that for $S_{1}$. More thorough informations are
provided by the distributions $P\left( k\right) $ and $P\left( l\right) $,
where $P\left( k\right) $ equips each node with the probability of accessing 
$k$ nodes and $P\left( l\right) $ corresponds to the probability associated
with minimal edge number $l$ between a node and the native state. In Fig.
1(a), the results of $P\left( k\right) $ are given with respect to different
move sets. In Fig. 1(b), the $P(l)$ curves develop very narrow widths of the
distributions for $S_{2}$ and $S_{3}$, sharply contrasting to the case for $%
S_{1}$. The results for $P\left( k\right) $ and $P\left( l\right) $ assure
us that, at high temperature, the kinetic features of $S_{2}$ and $S_{3}$
are similar but are drastically different from that of $S_{1}$. Meanwhile,
for finite temperatures, the structure peculiar to problem setting remains
important.

The method of Monte Carlo simulation is constrained by the condition of
detailed balance held on the transition between any of two connected nodes,
say from $c_{k}$ to $c_{m}$, 
\begin{equation}
P_{eq}\left( c_{k}\right) W\left( c_{k}\rightarrow c_{m}\right)
=P_{eq}\left( c_{m}\right) W\left( c_{m}\rightarrow c_{k}\right) ,
\label{eq1}
\end{equation}%
where $P_{eq}\left( c_{k}\right) $ is the equilibrium probability for the
conformation $c_{k}$, and the transition probability rate $W\left(
c_{k}\rightarrow c_{m}\right) $ can be factorized as 
\begin{equation}
W\left( c_{k}\rightarrow c_{m}\right) =p\left( c_{k}\rightarrow c_{m}\right)
\cdot P^{acc}\left( c_{k}\rightarrow c_{m}\right) ,  \label{eq2}
\end{equation}%
with the probability of applying the corresponding update $p\left(
c_{k}\rightarrow c_{m}\right) $ and the acceptance rate $P^{acc}\left(
c_{k}\rightarrow c_{m}\right) $. For the randomly chosen target node $c_{m}$
from the $k_{c_{k}}$ connected nodes, we have $p\left( c_{k}\rightarrow
c_{m}\right) =1/k_{c_{k}}$. While the acceptance ratio is $R=P^{acc}\left(
c_{k}\rightarrow c_{m}\right) /P^{acc}\left( c_{m}\rightarrow c_{k}\right) $%
, the condition of detailed balance of Eq. (\ref{eq1}) and the factorized
form of $W\left( c_{k}\rightarrow c_{m}\right) $ of Eq. (\ref{eq2}) lead to 
\begin{equation}
R=r_{k,m}\exp \left[ -\left( E_{c_{m}}-E_{c_{k}}\right) /T\right] ,
\label{eq3}
\end{equation}%
with $r_{k,m}=k_{c_{k}}/k_{c_{m}}$, where the Boltzmann constant is set to
unity and $T$ is the temperature. Consequently, 
\begin{equation}
P^{acc}\left( c_{k}\rightarrow c_{m}\right) =\frac{R}{1+R},  \label{eq4}
\end{equation}%
the acceptance rate depends on both the energy difference and the edge
numbers of two nodes. Therefore, our scheme bears interesting correspondence
to the embedding structure of the conformation-network. This is different
from the method suggested by Collet \cite{collet} who employed $p\left(
c_{k}\rightarrow c_{m}\right) =1/\left( 2M-5\right) $ but adopted $r_{k,m}=1$
for the acceptance ratio $R$ of Eq. (\ref{eq3}) above of the basis of the
move set $S_{1}$. Noteworthy is also the conventional scheme equipped with
randomly updating $c_{k}\rightarrow c_{m}$ and $r_{k,m}=1$ may violate the
condition of detailed balance\cite{collet}. Subsequently, the reliability as
well as the efficiency of the algorithm and the dynamic features with
respect to move sets are of concern.

Consider the standard deviation of probability distribution from the
equilibrium probability, $D\left( t\right) =(\sum_{k=1}^{K}\left(
P_{eq}\left( c_{k}\right) -\Pi \left( c_{k},t\right) \right) ^{2}/K)^{1/2}$,
where $\Pi \left( c_{k},t\right) $ is the occurrence probability of the
state $c_{k}$ during the time steps $t$. By generating the Monte Carlo
trajectories of $300$ billion steps at $T=2.50$ starting from the completely
extended chain, the values of $D\left( t\right) $ can be sampled at each $%
100,000$ Monte Carlo steps. The results depicted in Fig. 3 reveal
interesting issues. First of all, as seen from Fig. 2(a), Monte Carlo
simulations with randomly updating $c_{k}\rightarrow c_{m}$ and $r_{k,m}=1$,
which violates the detailed balance, cannot converge toward thermal
equilibrium\cite{collet}. Relative to this, employing the non-ergodic set $%
S_{1}$ plays minor role for the unachievable thermal equilibrium since only
one isolated node exists for $M=16$. In addition, the Collet's method, for
which one spend additional time in determining whether an update is applied
or not, tends to slow down the convergence toward the thermal equilibrium in
contrast to our new scheme. Meanwhile, Fig. 2(b) also shows up even more
rapid convergence for the move sets $S_{2}$ and $S_{3}$ by means of our new
scheme. But the choice of move sets appears to be insensitive to the
asymptotic behaviours of the standard deviation of probability distribution,
the corresponding scaling rule reads then $D\left( t\right) \sim t^{-1/2}$.
Thus, we conclude the reliability and efficiency for simulations based on
our scheme. The price to pay is that for each update the edge numbers of the
present and target nodes have to be known. \ \ 

For analyzing the perspectives of folding kinetics in relation to move sets,
we refer to the CTMC method proposed by Gillespie\cite{gillespie1}. The CTMC
method can be equipped with our scheme,\ and it suggests that the
probabibility of the transition $c_{k}\rightarrow c_{m}$ is 
\begin{equation}
P\left( c_{k}\rightarrow c_{m}\right) =\frac{W\left( c_{k}\rightarrow
c_{m}\right) }{W_{S}\left( c_{k}\right) }  \label{eq6}
\end{equation}%
and the probability of the occurrence of a transition from the current node $%
c_{k}$ at the time $\tau $ is 
\begin{equation}
P_{c_{k}}\left( \tau \right) =\left[ W_{S}\left( c_{k}\right) \right] \exp %
\left[ -\tau W_{S}\left( c_{k}\right) \right] ,  \label{eq7}
\end{equation}%
where $W\left( c_{k}\rightarrow c_{m}\right) $ is given by Eq. (\ref{eq2})
and $W_{S}\left( c_{k}\right) =\sum_{j=1}^{k_{c_{k}}}W\left(
c_{k}\rightarrow c_{j}\right) $. Accordingly, we update the node $c_{k}$ and
determine its time duration $\tau $ by using two random numbers $r_{1}$ and $%
r_{2}$ ranging between $0$ and $1$. With the transition probability of Eq. (%
\ref{eq6}), we calibrate the transition from $c_{k}$ to $c_{l}$ for $%
X_{l-1}<r_{1}\leq X_{l}$ subject to $X_{m}=\sum_{i=1}^{m}W\left(
c_{k}\rightarrow c_{i}\right) /W_{S}\left( c_{k}\right) $. Then, based on
Eq. (\ref{eq7}) we estimate the time duration $\tau $ at $c_{k}$ by setting $%
r_{2}=\exp \left[ -\tau W_{S}\left( c_{k}\right) \right] $ to yield $\tau
=-\ln \left( r_{2}\right) /W_{S}\left( c_{k}\right) $.

We first employ the CTMC method described in the last paragraph to measure
the first passage time $t_{f}$, defined as the required time from a
stretched chain to the first arrival of native state. The mean values, $%
\left\langle t_{f}\right\rangle $, at various temperatures for different
move sets are obtained by averaging over the results of $1000$ simulations.
The curves of $\log \left\langle t_{f}\right\rangle $ versus $1/T$ \ are
parabolic for all move sets. Thence, there exists the fastest folding
temperature $T_{\min }$ with the value $1.74$ for $S_{1}$, and $1.70$ for $%
S_{2}$ and $S_{3}$. The corresponding $\left\langle t_{f}\right\rangle $ at $%
T_{\min }$ is $1.2E5$ for $S_{1}$, $3.5E4$ for $S_{2}$, and $2.8E4$ for $%
S_{3}$. Moreover, there exists a linear behaviour for $\log \left\langle
t_{f}\right\rangle $ against $1/T$ at the region of low temperatures, and
this gives rise to the Arrhenius law as $\left\langle t_{f}\right\rangle
\sim \exp (\left\langle E_{b}\right\rangle /T)$ with the mean activation
energy $\left\langle E_{b}\right\rangle =6.72$ for $S_{1}$, $5.04$ for $%
S_{2} $, and $4.44$ for $S_{3}$. All these results are consistent with the
geometric features of the conformation-networks, namely that the network
associated with $S_{2}$ or $S_{3}$ owns a larger value of the mean number of
edges associated with a node and a smaller value of the mean minimal edge
number between a node and the native state in comparison with those for $%
S_{1}$.

The kinetic traps, referred to the states strongly prohibiting the folding
into the native one but locating at only few steps away from it, are
important for understanding the folding kinetics. Their identifications
often rely on the distinction of local energy minima without taking account
of the entropy effect. To make the criterion more effectively, we introduce
the measure, the average time duration of a state, in the frame of the CTMC
method to distinguish obstacle states. The average time duration of a state $%
c_{k}$ is, from Eq. (\ref{eq7}), given as $\left\langle \tau
_{c_{k}}\right\rangle =\int_{0}^{\infty }\tau P_{c_{k}}\left( \tau \right)
d\tau $, and it yields $\left\langle \tau _{c_{k}}\right\rangle
=1/W_{S}\left( c_{k}\right) $. By defining the ratio of $\left\langle \tau
_{c_{k}}\right\rangle $ to the average value of all states, $\eta
_{c_{k}}=\left\langle \tau _{c_{k}}\right\rangle /\tau _{ave}$ with $\tau
_{ave}=\sum_{k=1}^{K}\left\langle \tau _{c_{k}}\right\rangle /K$, as the
viscosity-factor associated with the state $c_{k}$, we can signify an
obstacle state of foldings by means of $\eta >1$.

The root mean square deviation (RMSD) of the location of a node $c_{k}$ from
that of the native state $c_{0}$ is defined as $d_{RMSD}\left( c_{k}\right) =%
\left[ \sum_{i=1}^{M}(\overrightarrow{r}_{i}\left( c_{k}\right) -%
\overrightarrow{r}_{i}\left( c_{0}\right) )^{2}/M\right] ^{1/2}$, where $%
\overrightarrow{r}_{i}\left( c_{k}\right) $ is the position vector for the $%
i $th monomer of the state $c_{k}$. The $d_{RMSD}\left( c_{k}\right) $ is
adopted as the conformational distance of $c_{k}$ from the native state.
Note that the other alternative is to take the minimal number of edges
between two nodes as the conformational distance\cite{chan}, but this
suffers from the ambiguity owing to the choice of move sets. By taking the
average of the $\eta $ values, $\left\langle \eta \right\rangle $, of the
states associated with the given $d_{RMSD}$ as the viscosity-factor of the $%
d_{RMSD}$, we show the results of $\left\langle \eta \right\rangle $ versus $%
d_{RMSD}$ at the fastest folding temperatures $T_{\min }$ of different move
sets in Fig. 3. The results indicate that the folding process can be divided
into two stages. One is the folding from the denatured to the semicompact
states for which, it corresponds to $d_{RMSD}\gtrsim 2$ and the
viscosity-factor variations among different move sets are very small. Owing
to the large scale change caused by rigid rotation, the folding time for $%
S_{2}$ and $S_{3}$ can be expected to be much shorter than that for $S_{1}$.
The other stage is the adjustment to the native from the semicompact states
corresponding to the range $d_{RMSD}\lesssim 2$. Because of the high $%
\left\langle \eta \right\rangle $ values, the searching of the native state
is a very slow process for all move sets. For a given $d_{RMSD}\lesssim 2$,
the inequality $\left\langle \eta \right\rangle _{S_{2}}>\left\langle \eta
\right\rangle _{S_{3}}>\left\langle \eta \right\rangle _{S_{1}}$ indicates
that the local moves are more effective for the searching. The strongest
obstacles of foldings for different move sets all occur at $d_{RMSD}\simeq
0.5$ which contains $9$ conformations. Among these, the state with the
largest $\eta $ value, shown in the inset of Fig. 3, is the same for
different move sets, and the corresponding $\eta $ value is $21.4$ for $%
S_{1} $, $50.9$ for $S_{2}$, and $37.6$ for $S_{3}$. From the numerical
results of other sequences, we notice that the strongest obstacle is not
necessary the same for different move sets, it may also vary with the
sequence; but the inequality $\eta _{S_{2}}>\eta _{S_{3}}>\eta _{S_{1}}$ for
the strongest obstacle of a sequence is always true. \ \ 

In summary, we present a detailed study on the methodology of lattice Monte
Carlo method, including the implementation of the algorithm and the move set
adopted in the simulations. A new method of implementing the Metropolis
algorithm, which is shown to satisfy the condition of detailed balance, is
proposed, and the characteristic features of different move sets in the
folding kinetics are given. In particular, we combine the CTMC method with
the new implementation to introduce a more effective quantity, called
viscosity-factor, to identify the kinetic traps.

This work was partially supported by the National Science Council of
Republic of China (Taiwan) under the Grant No. NSC 93-2212-M-033-005.

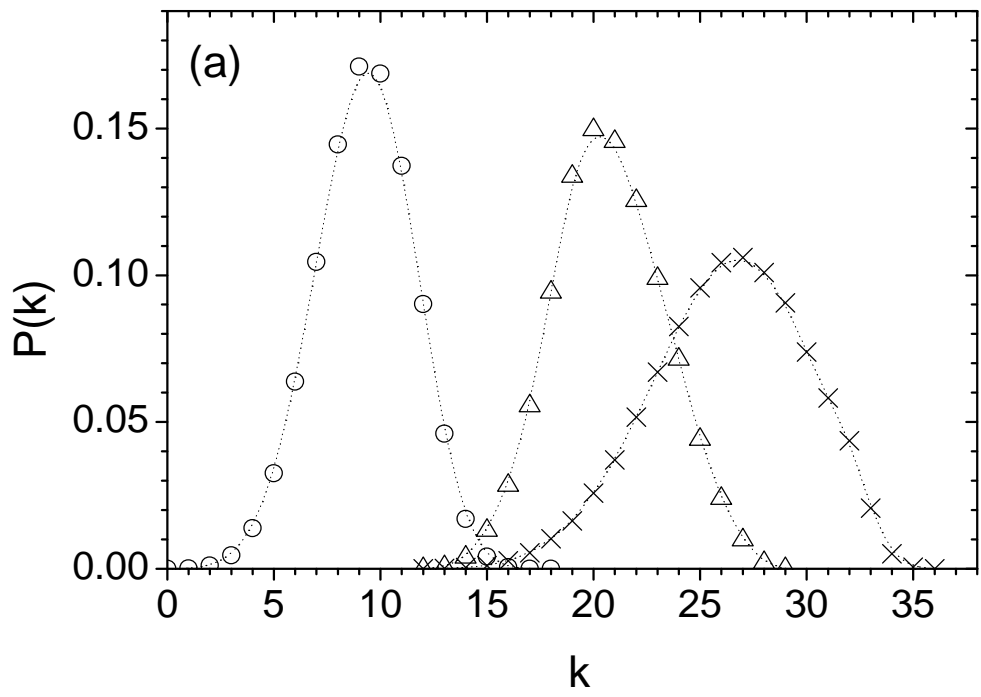
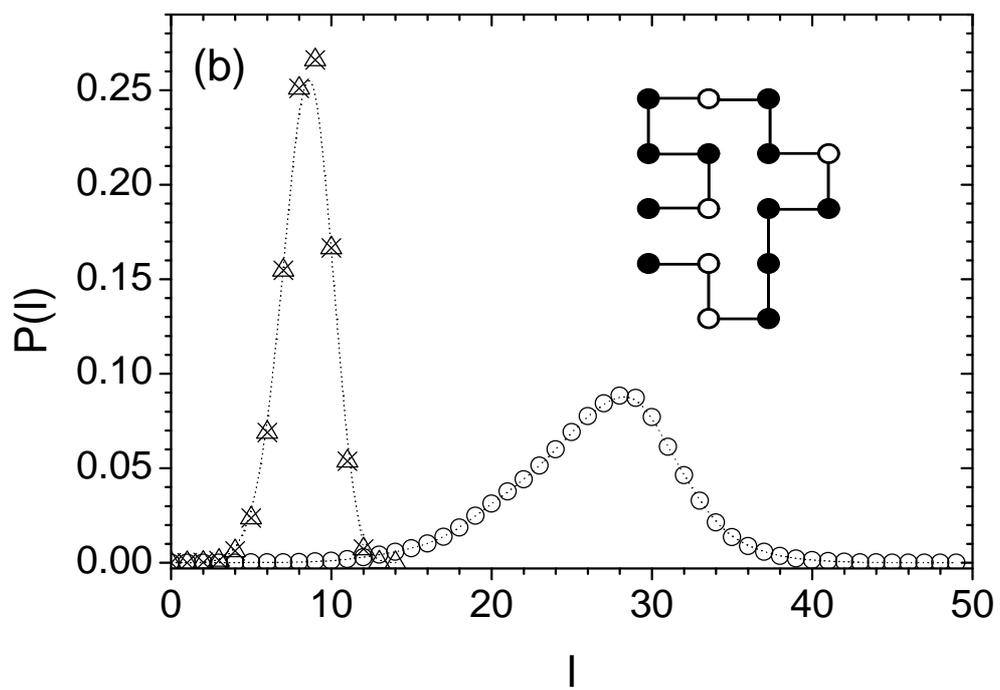
Fig. 1. $\left( a\right) $ The distribution function of the edge number
associated with a node, $P\left( k\right) $, versus the edge number, $k$;
and $\left( b\right) $ the distribution function of the distance $l$ from a
conformation to the native conformation, $P\left( l\right) $, versus the
distance $l$ for the move sets, $S_{1}$ (circles), $S_{2}$ (triangles), and $%
S_{3}$ (crosses). The inset shown in $\left( b\right) $ is the native
conformation of the model sequence.

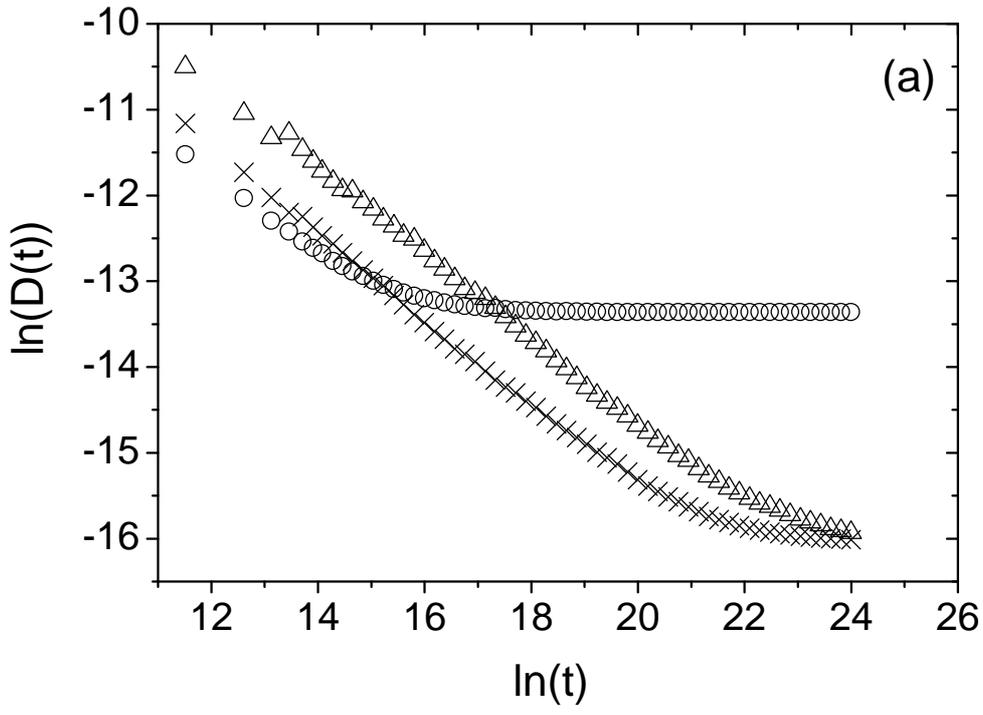
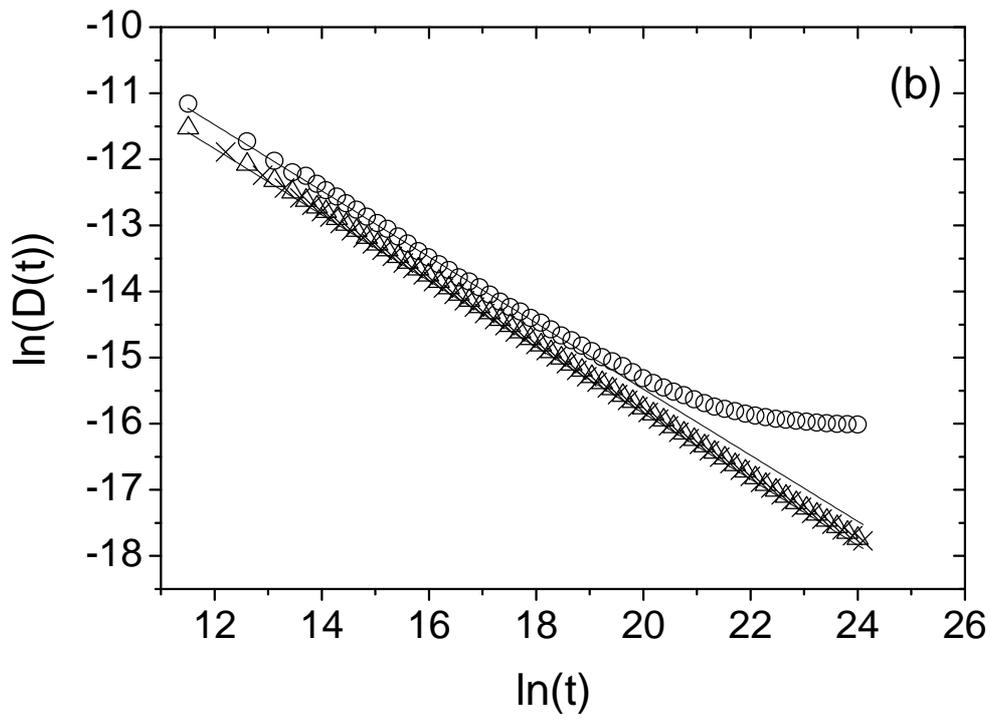
Fig. 2. The standard deviation of probability distribution from the
equilibrium probability, $D\left( t\right) $, as a function of the number of
Monte Carlo steps $t$, at $T=2.50$: $\left( a\right) $ the results of
conventional implementation (circles), Collet's method (triangles), and the
new proposed method (crosses) with the move set $S_{1}$; and $\left(
b\right) $ the results of the new proposed method with the move sets $S_{1}$%
(circles), $S_{2}$ (triangles), and $S_{3}$(crosses) with the straight lines
given by $D\left( t\right) \sim t^{-1/2}$. \ \ \ \ 

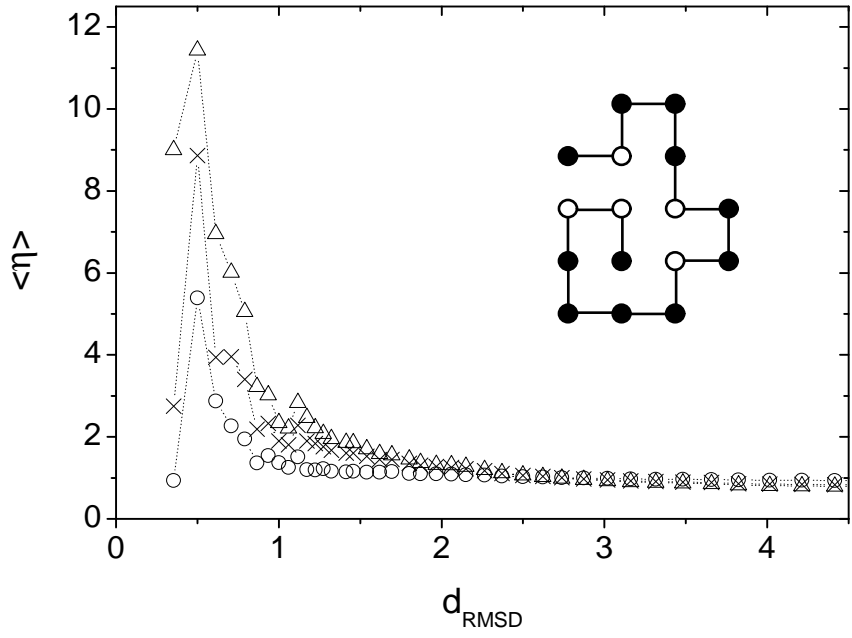
Fig. 3. The average value of viscosity-factors, $\left\langle \eta
\right\rangle $, versus the distance to the native conformation, $d_{RMSD}$,
at the fastest folding temperatures $T_{\min }$ for the move sets $S_{1}$%
(circles), $S_{2}$ (triangles), and $S_{3}$(crosses). The inset shows the
conformation with the largest $\eta $ value for the peak in the curve of $%
\left\langle \eta \right\rangle $ versus $d_{RMSD}$.

\end{document}